\documentclass[11pt,twoside]{amsart}

\usepackage[hmargin={3cm,3cm},vmargin={3cm,3cm},includehead]{geometry}
\usepackage{cite}
\usepackage{amsmath,amssymb}
\usepackage{amsfonts}
\usepackage{mathrsfs}

\newtheorem{stat}{Statement}

\newcommand{\ds}{\displaystyle}
\newcommand{\tensprod}{\mathop{\textrm{\Large $\otimes$}}}
\def\EXP{\textrm{{\large e}}}

\newcommand{\bosn}{\textrm{\scriptsize $\boldsymbol{N}$}}

\newcommand{\bos}{\boldsymbol{a}}

\newcommand{\kop}{\boldsymbol{k}}

\newcommand{\alg}{\mathcal{A}}
\newcommand{\pt}{\boldsymbol{n}}
\newcommand{\e}{\boldsymbol{e}}
\newcommand{\radius}{\boldsymbol{r}}

\newcommand{\rop}{\mathsf{R}}
\newcommand{\wop}{\boldsymbol{w}}
\newcommand{\ii}{\mathsf{i}}

\begin{document}

\title[Ground states of evolution operator]
{Ground states of Heisenberg evolution operator in discrete
three-dimensional space-time and
quantum discrete BKP equations}%

\author{Sergey M. Sergeev}

\address{Faculty of Informational Sciences and Engineering,
University of Canberra, Bruce ACT 2601}

\email{sergey.sergeev@canberra.edu.au}



\begin{abstract}
In this paper we consider three-dimensional quantum $q$-oscillator
field theory without spectral parameters. We construct an
essentially big set of eigenstates of evolution with unity
eigenvalue of discrete time evolution operator. All these
eigenstates belong to a subspace of total Hilbert space where an
action of evolution operator can be identified with quantized
discrete BKP equations (synonym Miwa equations). The key
ingredients of our construction are specific eigenstates of a
single three-dimensional $R$-matrix. These eigenstates are
boundary states for hidden three-dimensional structures of
$\mathscr{U}_q(B_n^{(1)})$ and $\mathscr{U}_q(D_n^{(1)})$.
\end{abstract}

\maketitle

\section*{Introduction}

Quantum $q$-oscillator system \cite{BS05,circular} in
three-dimensional space-time is the result of canonical
quantization of a Hamiltonian form \cite{Melbourne} of discrete
three-wave equations
\cite{Korepanov:1995,BogdanovKonopelchenko,DoliwaSantini}.

In the most general form \cite{Korepanov:1995,cft} the discrete
three-wave equations involve some extra parameters (spectral
parameters in quantum world) and correspond to a generic AKP-type
hierarchy of integrable systems. There are two special choices of
spectral parameters corresponding to the discrete-differential
geometry \cite{Sauer,BoSurBook} of discrete conjugate nets (syn.
quadrilateral nets) \cite{DoliwaSantini,BobenkoPinkall,KS98} --
either circular nets (syn. orthogonal nets) in Euclidean space or
ortho-chronous hyperbolic nets in Minkowski space. There are a lot
of equations associated with discrete nets, here we mean equations
for angular data (rotation coefficients) \cite{KS98,circular}.
Algebraically, circular and hyperbolic nets are distinguished by a
signature of determinant of rotation matrix.

For the latter case of hyperbolic nets the equations of motion
admit two constrains reducing a number of degrees of freedom of
Cauchy problem twice. One constraint corresponds to discrete BKP
equations (syn. Miwa equations) \cite{Hirota-Miwa}. Discrete BKP
equations appear in discrete differential geometry in many ways
\cite{KonopelchenkoSchiefBKP}, constraint for hyperbolic net just
clearly shows the reduction. The other constraint is a real form
on equations of motion (curiously, we discuss in fact six-wave
equations, they become three-wave upon this reality condition).

The discrete three-wave equations give a well posed Cauchy problem
in $2+1$ dimensional discrete space-time. The quantized discrete
three-wave equations are the Heisenberg equations of motion
defined by a discrete time evolution operator. The principal
question of quantum theory is the spectral problem for the
evolution operator (Schr\"odinger equation).

For arbitrary spectral parameters providing the unitarity of
evolution operator the spectral problem for it is rather
complicated. In this paper we study the evolution operator for
trivial spectral parameters corresponding in classics to the
hyperbolic net. We consider both the Fock space and modular
representations of $q$-oscillators. Quantum analogue of above
mentioned constraints provides a definition of a subspace of total
Hilbert space, and in this subspace we construct a big set
(presumably -- infinite set) of eigenstates with unity eigenvalue
of the evolution operator. Since we use in particular a quantum
analogue of dBKP constraint, we refer the resulting quantum theory
to as the quantum discrete BKP equations.

This paper is organized as follows. In Section 1 we formulate the
Cauchy problem in classics and quantum Heisenberg equation of
motion and give a formal definition of the evolution operator.
Section 1 is a brief outline of \cite{BS05,circular}. In Section 2
we discuss the reductions in classics and in quantum case. The
subspace of Hilbert space and eigenvectors of the evolution
operator are constructed in Section 3.

\section{$q$-oscillator field theory}

\subsection{Local Yang-Baxter and auxiliary Tetrahedron equations}

The most convenient form of auxiliary problem providing the
Hamiltonian equations of motion is the local Yang-Baxter equation.

Let $L_{\alpha\beta}[\alg]$ be a matrix acting in the tensor
product of two two-dimensional vector spaces $V_\alpha$ and
$V_\beta$,
\begin{equation}\label{L}
L_{\alpha,\beta}[\alg]\;=\;\left(\begin{array}{cccc} 1 & 0 & 0 & 0 \\
0 & \kop & \bos^+ & 0 \\ 0 & \bos^- & -\kop & 0 \\ 0 & 0 & 0 &
1\end{array}\right)\;,\quad \alg=(\kop,\bos^{\pm})\;.
\end{equation}
In classics, the fields $\alg$ are constrained by
\begin{equation}\label{constr}
\kop^2+\bos^+\bos^-=1
\end{equation}
and have the bracket
\begin{equation}\label{poisson}
\{\bos^+,\bos^-\}\;=\;\kop^2\;.
\end{equation}
Being quantized, $\alg$ is the $q$-oscillator algebra
\begin{equation}\label{qalg}
\kop\bos^{\pm}=q^{\pm 1} \bos^\pm \kop\;,\quad
\bos^+\bos^-=1-q^{-1}\kop^2\;,\quad \bos^-\bos^+=1-q\kop^2\;.
\end{equation}
The local Yang-Baxter equation is
\begin{equation}\label{lybe}
L_{\alpha,\beta}[\alg_1] L_{\alpha,\gamma}[\alg_2]
L_{\beta,\gamma}[\alg_3] \;=\; L_{\beta,\gamma}[\alg_3']
L_{\alpha,\gamma}[\alg_2'] L_{\alpha,\beta}[\alg_1']\;,
\end{equation}
it defines the following map
$\alg_1\times\alg_2\times\alg_3\to\alg_1'\times\alg_2'\times\alg_3'$:
\begin{equation}\label{map}
(\kop_2^{}\bos_1^\pm)'=\kop_3^{}\bos_1^\pm +
\kop_1^{}\bos_2^\pm\bos_3^\mp\;,\quad
(\bos_2^\pm)'=\bos_1^\pm\bos_2^\pm-\kop_1^{}\kop_3^{}\bos_2^\pm\;,\quad
(\kop_2^{}\bos_3^\pm)'=\kop_1^{}\bos_3^\pm +
\kop_3^{}\bos_1^\mp\bos_2^\pm\;,
\end{equation}
where in addition
\begin{equation}\label{kk}
(\kop_1\kop_2)'=\kop_1\kop_2\;,\quad
(\kop_2\kop_3)'=\kop_2\kop_3\;.
\end{equation}
These formulas work both in classics and in quantum case. In
classics, upon condition (\ref{constr}) for $\kop_j'$, map
(\ref{map}) preserves the symplectic structure (\ref{poisson}). In
quantum case map (\ref{map}) is automorphism of tensor cube of
$q$-oscillator algebra (\ref{qalg}) and therefore for any
irreducible representation of (\ref{qalg}) there exists an
operator $\rop_{123}$ such that
\begin{equation}\label{rmat}
\alg_j'=\rop_{123}^{}\alg_j^{}\rop_{123}^{-1}\;, \quad j=1,2,3,
\end{equation}
and the local Yang-Baxter equation becomes the auxiliary
tetrahedron equation
\begin{equation}\label{te}
L_{\alpha,\beta}[\alg_1] L_{\alpha,\gamma}[\alg_2]
L_{\beta,\gamma}[\alg_3] \rop_{123} \;=\; \rop_{123}
L_{\beta,\gamma}[\alg_3] L_{\alpha,\gamma}[\alg_2]
L_{\alpha,\beta}[\alg_1]\;.
\end{equation}
The key feature of map (\ref{map}) is that it is the square root
of unity,
\begin{equation}
\rop_{123}^2\alg_j\rop_{123}^{-2}\equiv \alg_j\;,
\end{equation}
and thus we are able to choose the overall normalization of
$\rop_{123}$ such that
\begin{equation}
\rop_{123}^2=1\;.
\end{equation}

Matrices $L$, eq. (\ref{L}), satisfy a free-fermions condition.
The local Yang-Baxter equation is the free-fermions form of
Korepanov zero curvature representation
\cite{Korepanov:1995,Korepanov:1999tmp} for the matrices of
auxiliary linear problem (linear problem for discrete three-wave
equations)
\begin{equation}
X_{\alpha\beta}[\alg]=\left(\begin{array}{cc} \kop & \bos^+ \\
\bos^- & -\kop \end{array}\right)\;.
\end{equation}
In the discrete differential geometry this is a matrix of rotation
coefficients, the case of hyperbolic net in Minkowski space is
fixed by condition $\det X = -1$.

\subsection{Lattice equations of motion}

Consider now a three-dimensional cubic lattice with basis vectors
$\e_1,\e_2,\e_3$,
\begin{equation}
\pt\;=\;n_1\e_1+n_2\e_2+n_3\e_3\;,\quad
n_1,n_2,n_3\in\mathbb{Z}\;.
\end{equation}
Map (\ref{map}) is the local form of equations of motion:
\begin{equation}
\alg_j^{}=\alg_{j,\pt}^{}\;\Rightarrow\;
\alg_j'=\alg_{j,\pt+\e_j}^{}\;,\quad j=1,2,3\;.
\end{equation}
With the space-time argument $\pt$, equation (\ref{lybe}) is
\begin{equation}\label{lybe-n}
L_{\alpha,\beta}[\alg_{1,\pt}] L_{\alpha,\gamma}[\alg_{2,\pt}]
L_{\beta,\gamma}[\alg_{3,\pt}] \;=\;
L_{\beta,\gamma}[\alg_{3,\pt+\e_3}]
L_{\alpha,\gamma}[\alg_{2,\pt+\e_2}]
L_{\alpha,\beta}[\alg_{1,\pt+\e_1}]\;,
\end{equation}
and equations (\ref{map}) become
\begin{equation}\label{eom}
\begin{array}{l}
\ds
\kop_{2,\pt+\e_2}^{}\bos_{1,\pt+e_1}^\pm=\kop_{3,\pt}^{}\bos_{1,\pt}^\pm
+ \kop_{1,\pt}^{}\bos_{2,\pt}^\pm\bos_{3,\pt}^\mp\;,\\
[3mm]
\ds
\bos_{2,\pt+\e_2}^\pm=\bos_{1,\pt}^\pm\bos_{2,\pt}^\pm-\kop_{1,\pt}^{}\kop_{3,\pt}^{}\bos_{2,\pt}^\pm\;,\\
[3mm]
\ds
\kop_{2,\pt+\e_2}^{}\bos_{3,\pt+\e_3}^\pm=\kop_{1,\pt}^{}\bos_{3,\pt}^\pm
+ \kop_{3,\pt}^{}\bos_{1,\pt}^\mp\bos_{2,\pt}^\pm\;,
\end{array}
\end{equation}
relation (\ref{kk}) provides in addition
\begin{equation}
\kop_{1,\pt+\e_1}\kop_{2,\pt+\e_2}=\kop_{1,\pt}\kop_{2,\pt}\;,\quad
\kop_{2,\pt+\e_2}\kop_{3,\pt+e_3}=\kop_{2,\pt}\kop_{3,\pt}\;.
\end{equation}

Since locally the equations of motion are given by the symplectic
map or by quantum automorphism, lattice equations (\ref{eom})
constitute classical Hamiltonian equations or quantum Heisenberg
evolution. The unique way to define the discrete time is
\begin{equation}
\tau\;=\;\tau(\pt)\;=\;n_1+n_2+n_3\;,
\end{equation}
so that all fields in the left hand side of (\ref{lybe-n})
correspond to time $\tau$ and all fields in the right hand side
correspond to one step forward time $\tau+1$. A choice of
space-like vectors is irrelevant. For instance, we can
choose\footnote{In general, any pair of $(n_1,n_2,n_3)$ can be
chosen as space-like coordinates. All choices are equivalent up to
translation operators.}
\begin{equation}
\e_\tau=\e_2\;,\quad \e_x=\e_1-\e_2\;,\quad \e_y=\e_3-\e_2\;,
\end{equation}
so that
\begin{equation}
\pt\;=\;\underbrace{n_1\e_x+n_3\e_y}_{\radius}+\tau\e_\tau\;,
\end{equation}
where $\radius$ stands for space-like position vector. This gives
\begin{equation}
\alg_{j,\pt}\;\equiv\;\alg_{j,\radius}(\tau)\;,
\end{equation}
and
\begin{equation}
\alg_{1,\pt+\e_1}=\alg_{1,\radius+\e_x}(\tau+1)\;,\quad
\alg_{2,\pt+\e_2}=\alg_{2,\radius}(\tau+1)\;,\quad
\alg_{3,\pt+\e_3}=\alg_{3,\radius+\e_y}(\tau+1)\;,
\end{equation}
so that equations (\ref{eom}) are precisely the discrete time
Hamiltonian flow.

Equations (\ref{eom}) is a well posed Cauchy problem for finite
size of constant time discrete surface with periodical boundary
conditions:
\begin{equation}
\radius\;=\;n_1\e_x+n_3\e_y\;,\quad n_1,n_3\in\mathbb{Z}_N
\end{equation}
In the quantum case, the Heisenberg equations of motion are
defined by the evolution operator,
\begin{equation}
\Phi(\tau+1)\;=\;U\;\Phi(\tau)\;U^{-1}\;,\quad U=\exp(\ii H)\;.
\end{equation}
In the form of intertwining (\ref{lybe-n}) the evolution operator
is defined by
\begin{equation}\label{lybe-ev}
L_{\alpha,\beta}[\alg_{1,\radius}]
L_{\alpha,\gamma}[\alg_{2,\radius}]
L_{\beta,\gamma}[\alg_{3,\radius}] \;=\;
U\;L_{\beta,\gamma}[\alg_{3,\radius+\e_y}]
L_{\alpha,\gamma}[\alg_{2,\radius}]
L_{\alpha,\beta}[\alg_{1,\radius+\e_x}]\;U^{-1}\;,
\end{equation}

Matrix element (or kernel) of the evolution operator can be
expressed in terms of matrix elements of $R$-matrix (\ref{rmat}).
Let $|\sigma'\rangle$ and $\langle \sigma|$ denote conjugated
bases in representation space of $q$-oscillator,
$\ds\sum|\sigma\rangle \langle \sigma|=1$ or $\ds
\int|\sigma\rangle \langle \sigma|=1$. Then $R$-matrix is defined
by its matrix element or kernel
\begin{equation}
\langle
\sigma_1^{},\sigma_2^{},\sigma_3^{}|\rop|\sigma_1',\sigma_2',\sigma_3'\rangle\;.
\end{equation}
Explicit form of matrix elements for Fock space representation and
kernel for modular representation can be found in Appendix. Matrix
element or kernel of the evolution operator is then evidently
\begin{equation}\label{evolution}
\langle\boldsymbol{\sigma}|U|\boldsymbol{\sigma}'\rangle \;=\;
\prod_{\radius\in\mathbb{Z}_N^2}\; \langle
\sigma_{1,\radius}^{},\sigma_{2,\radius}^{},\sigma_{3,\radius}^{}|\rop|
\sigma_{1,\radius+\e_x}',\sigma_{2,\radius}',\sigma_{3,\radius+\e_y}'\rangle\;.
\end{equation}
The evolution operator is unitary when $\rop$ is unitary. The
local structure of evolution operator corresponds to relativistic
casuality, thus we have the relativistic quantum field theory.

A complete set of integrals of motion is produced by an auxiliary
layer-to-layer transfer matrix. It is defined as follows:
\begin{equation}\label{TM}
T(x,y)\;=\;\mathop{\textrm{Trace}}_{V_{\boldsymbol{\alpha}}\times
V_{\boldsymbol{\beta}}\times V_{\boldsymbol{\gamma}}}\left(
D_{\boldsymbol{\alpha}}(x)D_{\boldsymbol{\beta}}(xy)
D_{\boldsymbol{\gamma}}(y) \prod_{n_1}^{\curvearrowright}
\prod_{n_3}^{\curvearrowleft}
L_{\alpha_{n_3}\beta_{n_2}}[\alg_{1,\radius}]
L_{\alpha_{n_3}\gamma_{n_1}}[\alg_{2,\radius}]
L_{\beta_{n_2}\gamma_{n_1}}[\alg_{3,\radius}]\right).
\end{equation}
Here we consider the tensor product of two-dimensional spaces,
\begin{equation}
V_{\boldsymbol{\alpha}}\;=\;\tensprod_{n_3\in\mathbb{Z}_N}
V_{\alpha_{n_3}}\;,\quad
V_{\boldsymbol{\beta}}\;=\;\tensprod_{n_2\in\mathbb{Z}_N}
V_{\beta_{n_2}}\;,\quad
V_{\boldsymbol{\gamma}}\;=\;\tensprod_{n_1\in\mathbb{Z}_N}
V_{\gamma_{n_1}}\;,
\end{equation}
so that matrix $L_{\alpha_{n_3},\beta_{n_2}}$ corresponds to the
components $V_{\alpha_{n_3}}\times V_{\beta_{n_2}}$, etc. The
ordered product in (\ref{TM}) is taken over $n_1$,
\begin{equation}
\prod_{n_1}^{\curvearrowright} f_{n_1}\;\stackrel{def}{=}\; f_0
f_1 f_2 \dots f_{N-1}\;,
\end{equation}
and over $n_3$,
\begin{equation}
\prod_{n_3}^{\curvearrowleft} f_{n_3}\;\stackrel{def}{=}\; f_{N-1}
f_{N-2}\dots f_1 f_0\;.
\end{equation}
Index $n_2$ of $V_{\beta}$ spaces is related to $n_1$ and $n_3$,
\begin{equation}
n_2=-n_1-n_3\quad (\tau=0)\;.
\end{equation}
Boundary matrices $D$ are defined by
\begin{equation}
D_{\boldsymbol{\alpha}}(x)\;=\;\tensprod_{n_3\in\mathbb{Z}_N}
D_{\alpha_{n_3}}(x)\;,\quad D_\alpha(x)=\left(\begin{array}{cc} 1
& 0 \\ 0 &
u\end{array}\right)\;\in\;\textrm{End}(V_\alpha)\;,\quad
\textrm{etc.}
\end{equation}
In the definition of ordered products $\mathbb{Z}_N^2$ invariance
is broken, however the final trace over all auxiliary spaces
restores $\mathbb{Z}_N^2$ invariance of transfer matrix
(\ref{TM}). Due to the ordering of products, the same
transfer-matrix can be identically rewritten as
\begin{equation}\label{TM-2}
\begin{array}{l}
\ds
T(x,y)\;=\;\\
[3mm]
\ds \mathop{\textrm{Trace}}_{V_{\boldsymbol{\alpha}}\times
V_{\boldsymbol{\beta}}\times V_{\boldsymbol{\gamma}}}\left(
D_{\boldsymbol{\alpha}}(x)D_{\boldsymbol{\beta}}(xy)
D_{\boldsymbol{\gamma}}(y) \prod_{n_1}^{\curvearrowright}
\prod_{n_3}^{\curvearrowleft}
L_{\beta_{n_2}\gamma_{n_1}}[\alg_{3,\radius+\e_y}]
L_{\alpha_{n_3}\gamma_{n_1}}[\alg_{2,\radius}]
L_{\alpha_{n_3}\beta_{n_2}}[\alg_{1,\radius+\e_x}] \right).
\end{array}
\end{equation}
where $n_2=-n_1-n_3-1$.

Comparing now the definition of evolution operator (\ref{lybe-ev})
and equivalence of (\ref{TM}) and (\ref{TM-2}), we deduce
\begin{equation}
U\;T(x,y)\;=\;T(x,y)\;U\;,
\end{equation}
i.e. the layer-to-later transfer matrix generates the invariants
of evolution,
\begin{equation}\label{TM-dec}
T(x,y)\;=\;\sum_{a,b} x^a y^b T_{a,b}\;,\quad 0\leq a,b\leq
2N\;,\quad |a-b|\leq N\;.
\end{equation}
From the theory of fermionic tetrahedron equations, see
\cite{BS05,supertetrahedron}, we know that the layer-to-layer
transfer matrices commute, i.e. the set of $T_{a,b}$ constitute a
family of $3N^2$ independent commutative operators (in classics --
quantities in involution)  -- the integrals of evolution.
Moreover, in classics the following equation
\begin{equation}
J(x,y)\;\stackrel{def}{=}\;\sum_{a,b} (-)^{a+b+ab} x^a y^b
T_{a,b}\;=\;0
\end{equation}
defines the spectral curve with genus $g\leq 3N^2-3N+1$ for the
evolution map \cite{Korepanov:1995}.

The constant-time section of three-dimensional cubic lattice is
known as the kagome lattice. Operator (\ref{TM}) is the
layer-to-layer transfer matrix on kagome lattice. The evolution
can be seen as a simultaneous shift of all $\beta$-lines on the
kagome lattice \cite{Korepanov:1995,Korepanov:1999tmp}.

\section{Constraints}

\subsection{Classical field theory}
There are two selected constraints for general
$\alg=(\kop,\bos^{\pm})$, $\kop^2\equiv 1-\bos^+\bos^-$, breaking
the Hamiltonian structure (\ref{poisson}) but preserved by map
(\ref{map}) and therefore by the equations of motion (\ref{eom}).
They are:
\begin{equation}\label{B}
B\;:\quad \bos^+=1-\kop\;,\quad \bos^-=1+\kop
\end{equation}
and
\begin{equation}\label{D}
D\;:\quad \bos^+=\bos^-\;.
\end{equation}
Constraint ``$B$'' results the map
\begin{equation}\label{B-map}
\kop_1'=\frac{\kop_1\kop_2}{\kop_1+\kop_3-\kop_1\kop_2\kop_3}\;,\quad
\kop_2'=\kop_1+\kop_3-\kop_1\kop_2\kop_3\;,\quad
\kop_3'=\frac{\kop_2\kop_3}{\kop_1+\kop_3-\kop_1\kop_2\kop_3}
\end{equation}
This is the well known representation of discrete BKP equation
\cite{Hirota-Miwa,KonopelchenkoSchiefBKP} as the map satisfying
the functional tetrahedron equation
\cite{Kashaev:1995ky,KKS:1998}. The substitution
\begin{equation}
\kop_{1,\pt}=u\frac{\tau_{\pt+\e_2}\tau_{\pt+\e_3}}{\tau_{\pt}\tau_{\pt+\e_2+\e_3}}\;,\quad
\kop_{2,\pt}=v\frac{\tau_{\pt}\tau_{\pt+\e_1+\e_3}}{\tau_{\pt+\e_1}\tau_{\pt+\e_3}}\;,\quad
\kop_{3,\pt}=w\frac{\tau_{\pt+\e_1}\tau_{\pt+\e_2}}{\tau_{\pt}\tau_{\pt+\e_1+\e_2}}
\end{equation}
converts equations of motion (\ref{eom}) into the four-term
bilinear Miwa equation:
\begin{equation}
v\tau_{\pt+\e_1+\e_2+\e_3}\tau_{\pt}\;=\;
u\tau_{\pt+\e_1+\e_2}\tau_{\pt+\e_3} + w
\tau_{\pt+\e_2+\e_3}\tau_{\pt+\e_1} -
uvw\tau_{\pt+\e_1+\e_3}\tau_{\pt+\e_2}\;.
\end{equation}
The ``$D$''-constraint (\ref{D}) is just a real form on equations
of motion. For the Cauchy problem both these constraints mean the
reduction of the number of degrees of freedom twice, $6N^2\to
3N^2$. Also, the number of independent invariants of evolution is
reduced nearly twice since
\begin{equation}\label{integrals-red}
B,D\;:\quad T_{a,b}=T_{2N-a,2N-b}\;.
\end{equation}

\subsection{Quantum constraints: Fock space representation}
Now we are back to quantum world and $q$-oscillator algebra
(\ref{qalg}). Here we consider the Fock space representation of
$q$-oscillators over the Fock vacuum $|0\rangle$,
\begin{equation}\label{Frep}
\bos^-|0\rangle =0\;,\quad |n\rangle \sim \bos^{+
n}|0\rangle\;,\quad \kop=q^{\bosn+1/2}\;,\quad \bosn|n\rangle =
|n\rangle n\;.
\end{equation}
Here $\bosn$ is the occupation number operator. If $0<q<1$ and
$(\bos^-)^\dagger = \bos^+$, the $R$-matrix and evolution
operators are unitary. Constraints $B$ and $D$, eqs.
(\ref{B},\ref{D}), are conditions for states:
\begin{equation}\label{q-B}
B\;:\quad u\bos^+|\psi^{B}(u)\rangle \;=\;
(1-q^{-1/2}\kop)|\psi^{B}(u)\rangle
\end{equation}
and
\begin{equation}\label{q-D}
D\;:\quad (\bos^--u\bos^+)|\psi^{D}(u)\rangle \;=\; 0\;.
\end{equation}
Parameter $u$ here is an extra useful $\mathbb{C}$-valued
parameter making norms of $|\psi^{B}(u)\rangle$ and
$|\psi^D(u)\rangle$ finite. Solutions to (\ref{q-B},\ref{q-D}) are
respectively
\begin{equation}\label{Bose-psivector-1}
|\psi^{B}(u)\rangle \;=\;\sum_{n=0}^\infty
\frac{(u\bos^+)^n}{(q;q)_n}|0\rangle
\;=\;(u\bos_j^+;q)_\infty^{-1}|0\rangle \;,
\end{equation}
and
\begin{equation}\label{Bose-psivector}
|\psi^{D}(u)\rangle \;=\;\sum_{n=0}^\infty
\frac{(u\bos_j^{+2})^n}{(q^4;q^4)_n}|0\rangle
\;=\;(u\bos_j^{+2};q^4)_\infty^{-1}|0\rangle\;.
\end{equation}
In these formulas we use Pochhammer symbol:
\begin{equation}\label{poh}
(x;p)_n=(1-x)(1-px)\dots (1-p^{n-1}x)\;.
\end{equation}
Norms of $|\psi^{B}\rangle$ and $|\psi^{D}\rangle$ are given by
\begin{equation}
\langle\psi^B(v)|\psi^B(u)\rangle =
\frac{(-qvu;q)_\infty}{(vu;q)_\infty}\;,\quad
\langle\psi^D(v)|\psi^D(u)\rangle =
\frac{(q^2vu;q^4)_\infty}{(vu;q^4)_\infty}\;.
\end{equation}
Note, ``$B$''-relation (\ref{q-B}) provides
\begin{equation}\label{q-B-2}
\bos^-|\psi^B(u)\rangle \;=\; u
(1+q^{1/2}\kop)|\psi^B(u)\rangle\;.
\end{equation}

\begin{stat}\label{st1} There are two types of invariant subspaces of $\rop$-matrix in the Fock space
representations,
\begin{equation}\label{eigen}
\rop_{123}|\Omega\rangle \;=\; |\Omega\rangle\;.
\end{equation}
They are
\begin{equation}\label{eigen-B}
|\Omega\rangle\;=\;|\Omega^B\rangle \;=\;
|\psi^B(u)\rangle_1\otimes |\psi^B(uv)\rangle_2\otimes
|\psi^B(v)\rangle_3
\end{equation}
and
\begin{equation}\label{eigen-D}
|\Omega\rangle \;=\;|\Omega^D\rangle \;=\;
|\psi^D(u)\rangle_1\otimes |\psi^D(uv)\rangle_2\otimes
|\psi^D(v)\rangle_3\;.
\end{equation}
for arbitrary $u$ and $v$.
\end{stat}

Since $|\psi^D\rangle$ involves only even occupation numbers,
$|\Omega^D\rangle$ is the eigenstate of
$\rop_{123}'\;=\;(-)^{\bosn_2}\rop_{123}^{}$ what corresponds to
Euclidean rotation coefficients $X=\left(\begin{array}{cc} \kop &
\bos^+ \\ -\bos^- & \kop\end{array}\right)$ with $\det X=1$.

In the resent paper \cite{D} the Reader can find a scenario how to
use the vectors $|\psi^D(u)\rangle$ as the three-dimensional
boundary states to reproduce the $R$-matrices, $L$-operators and
representation structure of $\mathscr{U}_q(D_n^{(1)})$. In a
similar and even more simple way the vectors $|\psi^B(u)\rangle$
can be used as the boundary states reproducing
$\mathscr{U}_q(B_n^{(1)})$. However, the quantum groups exercises
are not quite relevant to the study of three-dimensional evolution
operator.

Using formulas for map (\ref{map}), we can instantly obtain
\begin{equation}\label{q-BKP}
\rop_{123}\kop_2|\Omega^B\rangle \;=\;
(\kop_1+\kop_3-\kop_1\kop_2\kop_3+ (q^{1/2}-q^{-1/2})
\kop_1\kop_3)|\Omega^B\rangle\;,
\end{equation}
what is the quantum counterpart of (\ref{B-map}). However,
decomposition of
$\rop_{123}F(\kop_1,\kop_2,\kop_3)|\Omega^B\rangle$ for arbitrary
function $F$ is not well defined, and the modular representation
is more preferable.

\subsection{Quantum constraints: modular representation}

For the modular representation of $q$-oscillator we use the
Heisenberg pair $\sigma,p$,
\begin{equation}
[\sigma,p]\;=\;\frac{\ii}{2\pi}
\end{equation}
The oscillator is given by
\begin{equation}\label{mod-rep}
q=\EXP^{\ii\pi b^2}\;,\quad \kop=-\ii\EXP^{\pi b \sigma}\;,\quad
\wop=\EXP^{2\pi b p}\;,\quad \bos^{\pm}=(1-q^{\mp
1}\kop^2)^{1/2}\wop^{\pm 1}\;.
\end{equation}
The dual one is then
\begin{equation}\label{mod-rep-dual}
\overline{q}=\EXP^{-\ii\pi b^{-2}}\;,\quad
\overline{\kop}=\ii\EXP^{\pi b^{-1} \sigma}\;,\quad
\overline{\wop}=\EXP^{2\pi b^{-1} p}\;,\quad
\overline{\bos^\pm}=(1-\overline{q}^{\mp
1}\overline{\kop}^2)^{1/2}\overline{\wop}^{\mp 1}\;.
\end{equation}
Define a state $|\Phi\rangle$ by its wave function:
\begin{equation}\label{Phi}
\langle \sigma |\Phi\rangle \;=\; \Phi(\sigma)\;=\;
\exp\left(\frac{1}{8}\int_{\mathbb{R}_+} \frac{\EXP^{-2\ii\sigma
w}}{\sinh(bw) \cosh(b^{-1}w)} \frac{dw}{w}\right)\;.
\end{equation}
The modular invariance $b\leftrightarrow b^{-1}$ is broken. The
state $|\Phi\rangle$ satisfies
\begin{equation}
\bos^+|\Phi\rangle \;=\; (1-q^{-1/2}\kop^2)|\Phi\rangle\;,
\end{equation}
what is the $B$-type condition, and
\begin{equation}
(\overline{\bos^-}-\overline{\bos^+})|\Phi\rangle \;=\; 0\;,
\end{equation}
what is the $D$-type condition. The asymptotic of $\Phi$ is
\begin{equation}
\Phi(\sigma)_{\sigma\to -\infty}\to 1\;,\quad
\Phi(\sigma)_{\sigma\to +\infty}\to \EXP^{-\pi b^{-1}\sigma/2}\;.
\end{equation}
Let next
\begin{equation}
|\Phi_{d}\rangle \;=\; \kop^d|\Phi\rangle\;,
\end{equation}
where $d$ is real (and integer). A test of asymptotic of function
$\Phi(\sigma)$ shows that the state $|\Phi_{d}\rangle$ has a
finite norm if
\begin{equation}\label{convergency}
0<d<\frac{1}{2b^2}\;.
\end{equation}
Thus, in what follows we imply the quantum regime near
quasi-classical point $b=0$:
\begin{equation}
0< b\ll 1\;,
\end{equation}
so that $d$ can be reasonably high.

\begin{stat}\label{st2} For the modular representation of $q$-oscillators
operator $\rop_{123}$ has the eigenstates (\ref{eigen}) given by
\begin{equation}\label{eigen-M}
|\Omega\rangle \;=\;
|\Phi_{d}\rangle_1\otimes|\Phi_{d+d'}\rangle_2 \otimes
|\Phi_{d'}\rangle_3
\end{equation}
\end{stat}
Moreover, due to the asymptotic of $R$, one can verify that if a
state $F(\kop_1,\kop_2,\kop_3)|\Omega\rangle$ has a finite norm,
then the convolution
$\rop_{123}F(\kop_1,\kop_2,\kop_3)|\Omega\rangle$ is convergent
and therefore a map $F\to F'$ in the space of meromorphic
functions with proper asymptotic (\ref{convergency})
\begin{equation}\label{BKP-map}
\rop_{123}F(\kop_1,\kop_2,\kop_3)|\Omega\rangle\;=\;
F'(\kop_1,\kop_2,\kop_3)|\Omega\rangle
\end{equation}
is well defined. This extends equation (\ref{q-BKP}) to a subspace
of whole Hilbert space -- quantum BKP theory. In the classical
limit $b\to 0$ ($q\to 1$) this map becomes the rational one,
\begin{equation}
F'(\kop_1,\kop_2,\kop_3)\;\mathop{\rightarrow}_{b\to 0}\;
F(\kop_1',\kop_2',\kop_3')\;,
\end{equation}
where $\kop_j'$ are given by (\ref{B-map}).

\section{Ground states of evolution operator}

\subsection{Fock space representation}

The straightforward extensions of (\ref{eigen-B},\ref{eigen-D}) to
the whole constant time surface is
\begin{equation}
|\Omega\rangle \;=\; \prod_{\radius=n_1\e_x+n_3\e_y}
|\psi(u_{n_3})\rangle_{1,\radius}\otimes
|\psi(u_{n_3}v_{n_1})\rangle_{2,\radius}\otimes
|\psi(v_{n_1})\rangle_{3,\radius}
\end{equation}
for $B$ and $D$ states in Fock space. So defined $|\Omega\rangle$
are the eigenstates of evolution operator,
\begin{equation}
U\;|\Omega\rangle \;=\; |\Omega\rangle\;,
\end{equation}
for arbitrary $u_{n_3}$ and $v_{n_1}$. Series decomposition of
$|\Omega\rangle$ gives an infinite set of eigenstates
corresponding to fixed eigenvalues of
\begin{equation}\label{JK}
J_{n_3}\;=\;\prod_{n_1}\kop_{1,n_1\e_x+n_3\e_y}\kop_{2,n_1\e_x+n_3\e_y}\;,\quad
K_{n_1}\;=\;\prod_{n_3}\kop_{2,n_1\e_x+n_3\e_y}\kop_{3,n_1\e_x+n_3\e_y}\;.
\end{equation}
Set of operators $J$ and $K$ belong to the family of integrals of
motion (\ref{TM-dec}).

Remarkably, so constructed eigenstates of the evolution operator
are not in general eigenstates of all integrals of motion. Thus,
the more general form of evolution eigenstates is given by
$(u,v)$-decomposition of
\begin{equation}
\label{general} \prod_{a,b} T_{a,b}^{n_{a,b}} |\Omega\rangle,
\end{equation}
where $T_{a,b}$ defined by (\ref{TM-dec}). Recall that
$T_{a,b}|\Omega\rangle = T_{2N-a,2N-b}|\Omega\rangle$, in
(\ref{general}) we use a set of independent $T_{a,b}$ with
non-diagonal action on $|\Omega\rangle$.

\subsection{Modular representation}
For the modular representation the basic eigenstate of evolution
operator is given by
\begin{equation}\label{trivial}
|\Omega\rangle \;=\; \prod_{n_3} J_{n_3}^{d_{n_3}}
\;\cdot\;\prod_{n_1} K_{n_1}^{d_{n_1}'}\;\cdot \;\prod_{\radius,j}
|\Phi\rangle_{j,\radius}\;,
\end{equation}
where $J,K$ are given by (\ref{JK}) and $|\Phi\rangle$ is given by
(\ref{Phi}). This state has the finite norm if
\begin{equation}\label{conv-2}
0<d_{n_3}^{},d_{n_1}'<\frac{1}{2b^2}\;,
\end{equation}
confer with (\ref{convergency}). Extended set of eigenstates is
given by (\ref{general}). Note that $J,K$ pre-factor in
(\ref{trivial}) is an element of
$\ds\prod_{a,b}T_{a,b}^{n_{a,b}}$. For sufficiently small $b$ the
states (\ref{general}) have finite norms.

However, contrary to the Fock space case, the states
\begin{equation}
\prod_{a,b} \overline{T}_{a,b}^{n_{a,b}}|\Omega\rangle\;,
\end{equation}
where $\overline{T}_{a,b}$ are modular partners to $T_{a,b}$, do
not have finite norms.

\section{Conclusion}

General evolution operators for three-dimensional field theories
are given by (\ref{evolution}),
\begin{equation}\label{evolution2}
\langle\boldsymbol{\sigma}|U|\boldsymbol{\sigma}'\rangle \;=\;
\prod_{\radius\in\mathbb{Z}_N^2}\; \langle
\sigma_{1,\radius}^{},\sigma_{2,\radius}^{},\sigma_{3,\radius}^{}|\mathcal{R}|
\sigma_{1,\radius+\e_x}',\sigma_{2,\radius}',\sigma_{3,\radius+\e_y}'\rangle\;.
\end{equation}
where the constant $\rop$-matrix is replaced by
\begin{equation}\label{R-sp-fock}
\mathcal{R}_{123}\;=\;\varrho^{-1}\EXP^{\ii(\pi-\phi_2)\bosn_2}\;\rop_{123}\;\EXP^{-\ii\phi_1\bosn_1-\ii\phi_3\bosn_3}
\end{equation}
for the Fock space representation, $\bosn_j$ here are the
occupation numbers, and by
\begin{equation}\label{R-sp-mod}
\mathcal{R}_{123}\;=\;\varrho^{-1}\EXP^{2\ii\eta\phi_2\sigma_2}\;\rop_{123}\EXP^{-2\ii\eta\phi_1\sigma_1-2\ii\eta\phi_3\sigma_3}
\end{equation}
for the modular representation, $\eta$ here is the crossing
parameter, $\ds  \eta\;=\;\frac{1}{2}(b+b^{-1})$.

$R$-matrices and evolution operators are unitary for real spectral
parameters $\phi_i$. Quantum field theories have good
quasi-classical limits for positive $\phi_i$ corresponding to
sides of certain hyperbolic triangles \cite{cft}. Spectra of
evolution operators essentially depend on values of spectral
parameters. Presumably, the positiveness of spectral parameters
and proper choice of unitary normalization factor $\varrho$ in
(\ref{R-sp-fock},\ref{R-sp-mod}) provide a good physical
interpretation of the evolution spectra in terms of ground state
and elementary excitations.

In this paper we consider the special case of trivial spectral
parameters. The main result of the paper is the observation of
essential degeneracy of ground state $U=1$ of the spectral
parameters free case. The eigenstates constructed belong to a
subspace of Hilbert space where Heisenberg evolution is a
$q$-analogue of discrete BKP equations. These eigenstates however
are not orthogonal and do not solve the problem of diagonalization
of all integrals of motion.

\bigskip

\noindent\textbf{Acknowledgements.} I would like to thank  V.
Bazhanov, R. Kashaev, V. Mangazeev and P. Vassiliou for valuable
discussions and fruitful collaboration.


\def\cprime{$'$} \def\cprime{$'$}
\providecommand{\bysame}{\leavevmode\hbox
to3em{\hrulefill}\thinspace}
\providecommand{\MR}{\relax\ifhmode\unskip\space\fi MR }
\providecommand{\MRhref}[2]{%
  \href{http://www.ams.org/mathscinet-getitem?mr=#1}{#2}
} \providecommand{\href}[2]{#2}

\appendix

\section{Matrix elements and kernel}

\subsection{Matrix elements}
Matrix elements of $\rop$-matrix of Statement \ref{st1} in the
unitary Fock basis
\begin{equation}
F^+\;:\quad |n\rangle
\;=\;\frac{\bos^{+n}}{\sqrt{(q^2;q^2)_n}}|0\rangle\;,\quad n\geq 0
\end{equation}
are given by
\begin{equation}\label{R-fock}
\begin{array}{l}
\ds \langle n_1^{}n_2^{}n_3^{}|\rop|n_1'n_2'n_3'\rangle\;=\;
\delta_{n_1^{}+n_2^{},n_1'+n_2'} \delta_{n_2^{}+n_3^{},n_2'+n_3'}
\prod_{i=1}^3 c_{n_i^{},n_i'}\times\phantom{xxxxxxxxxx}
\\
\\
\phantom{xxxxxxxxxxxxxx}\ds q^{n_1^{}n_3^{}+n_2'}\;
\frac{1}{2\pi\ii} \oint \;\frac{dz}{z^{n_2'+1}}\;
\frac{(-q^{2+n_1'+n_3'}z;q^2)_\infty
(-q^{-n_1^{}-n_3^{}}z;q^2)_\infty}{(-q^{+n_1^{}-n_3^{}}z;q^2)_\infty(-q^{-n_1^{}+n_3^{}}z;q^2)_\infty}\;,
\end{array}
\end{equation}
where
\begin{equation}
c_{n,n'}=\sqrt{\frac{(q^2;q^2)_{n'}}{(q^2;q^2)_n}}\quad
\textrm{for}\quad n=0,1,2,3\dots
\end{equation}
Here Pochhammer's symbol and Euler's quantum dilogarithm are
defined by (\ref{poh}). Coefficients $c_{n,n'}$ are just gauge
factors. The \emph{clockwise} integration loop in (\ref{R-fock})
circles all poles from dilogarithms but not includes $z=0$. The
Cauchy integral expression is equivalent to generating functions
from \cite{supertetrahedron}

Formula (\ref{R-fock}) serves in fact eight different
$\rop$-matrices. The occupation numbers in (\ref{R-fock}) are in
general integers,
\begin{equation}
n\;\in\;\mathbb{Z}\;=\;\mathbb{Z}_{<0}\;\oplus\;\mathbb{Z}_{\geq
0}\;.
\end{equation}
This corresponds to the direct sum of Fock and anti-Fock
representations,
\begin{equation}
F\;=\;F^-\oplus F^+\;.
\end{equation}
Matrix (\ref{R-fock}) has the block-diagonal structure in
\begin{equation}
F_1^{\epsilon_1}\otimes F_2^{\epsilon_2}\otimes
F_3^{\epsilon_3}\;,\quad \epsilon_i=\pm\;.
\end{equation}
$\rop$-matrix is unitary in four blocks with
$\epsilon_1\epsilon_2\epsilon_3=+$. In anti-Fock components
quantum constraints (\ref{q-B},\ref{q-D}) should be slightly
modified. In this paper we use the block $F_1^+\otimes
F_2^+\otimes F_3^+$ where (\ref{R-fock}) is equivalent to constant
$\rop$-matrix from \cite{circular}.

\subsection{Kernel for modular representation}

The kernel of $\rop$-matrix of Statement \ref{st2} in
representation (\ref{mod-rep},\ref{mod-rep-dual}) is given by
\cite{circular}
\begin{equation}\label{R-mod}
\begin{array}{l}
\ds \langle
\sigma_1^{}\sigma_2^{}\sigma_3^{}|\rop|\sigma_1'\sigma_2'\sigma_3'\rangle\;=\;
\delta_{\sigma_1^{}+\sigma_2^{},\sigma_1'+\sigma_2'}
\delta_{\sigma_2^{}+\sigma_3^{},\sigma_2'+\sigma_3'}
\sqrt{\frac{\varphi(\sigma_1)\varphi(\sigma_2)\varphi(\sigma_3)}
{\varphi(\sigma_1')\varphi(\sigma_2')\varphi(\sigma_3')}}
\\
\\
\ds
\EXP^{-\ii\pi(\sigma_1^{}\sigma_3^{}-\ii\eta(\sigma_1^{}+\sigma_3^{}-\sigma_2'))}\;
\int_{\mathbb{R}} \;du\; \EXP^{2\pi\ii u (\sigma_2'-\ii\eta)}\;
\frac{\varphi(u+\frac{\sigma_1'+\sigma_3'+\ii\eta}{2})\varphi(u+\frac{-\sigma_1-\sigma_3+\ii\eta}{2})}{\varphi(u+\frac{\sigma_1-\sigma_3-\ii\eta}{2})\varphi(u+\frac{\sigma_3-\sigma_1-\ii\eta}{2})}
\end{array}
\end{equation}
where $\varphi(\sigma)$ is the Barns-Faddeev non-compact quantum
dilogarithm \cite{Faddeev:1995}
\begin{equation}
\varphi(z)\;=\;\exp\left(\ds \frac{1}{4}\int_{\mathbb{R}+\ii 0}
\frac{\EXP^{-2\ii zw}}{\textrm{sinh}(bw)\textrm{sinh}(w/b)}\
\frac{dw}{w}\right)\;.\label{fi-def}
\end{equation}
and $\eta=\frac{1}{2}(b+b^{-1})$ is the crossing-parameter. In
this paper we imply the regime of big crossing parameter, $0<b\ll
1$. Both quantum dilogarithm $\varphi(\sigma)$ and asymmetric
function $\Phi(\sigma)$ (\ref{Phi}) are analytical in the strip
\begin{equation}
-\eta < \textrm{Im}(\sigma) < \eta\;.
\end{equation}

\end{document}